\documentclass[12pt]{iopart}
\usepackage{cite}
\usepackage{color}
\usepackage{graphicx}
  \expandafter\let\csname equation*\endcsname\relax
  \expandafter\let\csname endequation*\endcsname\relax
\usepackage{amsmath}
\begin{document}
\title[Stochastic Wilson-Cowan networks with delay]{Stochastic Wilson-Cowan models 
of neuronal network dynamics with memory and delay}

\author{Igor Goychuk$^1$ and Andriy Goychuk$^2$ }

\address{$^1$ Institute for Physics and Astronomy, University of Potsdam, Karl-Liebknecht-Str. 24/25,  
14476 Potsdam-Golm, Germany}
\ead{igoychuk@uni-potsdam.de, corresponding author}

\address{$^2$ Faculty of Physics, Ludwig Maximilian University of Munich, 
Munich, Germany}
\ead{andriy.goychuk@me.com}

\begin{abstract}  We consider a simple Markovian class of the stochastic
Wilson-Cowan type models of neuronal network dynamics, which incorporates
stochastic delay caused by the existence of a  refractory period of neurons.
From the point of view of the dynamics of the individual elements, we are
dealing with a network of non-Markovian stochastic two-state oscillators
with memory which are coupled globally in a mean-field fashion. This
interrelation of a higher-dimensional Markovian and lower-dimensional 
non-Markovian dynamics is discussed in its relevance to the general problem
of the network dynamics of complex elements possessing memory. The simplest
model of this class is  provided by a three-state Markovian neuron with
one refractory state, which causes firing delay with an exponentially decaying
memory within the two-state reduced model. This basic model is used to study 
critical avalanche dynamics (the noise sustained criticality) in a balanced 
feedforward network consisting of the excitatory and inhibitory neurons.
Such avalanches emerge due to the network size dependent noise (mesoscopic noise). 
Numerical simulations reveal an intermediate power law in the distribution of
avalanche  sizes with the critical exponent around $-1.16$. 
We show that
this power law is robust upon a variation of the refractory time over
several orders of magnitude. However, the
avalanche time distribution is biexponential. It does not reflect any genuine power
law dependence. 

\end{abstract}

\pacs{05.40.-a,87.19.lc,87.19.lj,87.19.ll}
\submitto{}
%\maketitle

\section{Introduction}

Network complexity pervades biology and medicine \cite{Sole}, 
and  the human organism
can be considered as an integrated complex network of different physiological systems \cite{Bashan} such
as circulatory and respiratory systems, visual system, digestive and endocrine systems, etc., 
which are coordinated by 
autonomic and central nervous systems including the brain.
The dynamics of the sleep-wake transitions during the 
sleep of humans and other mammals \cite{Lo1,Lo2,Lo3} presents one of 
important examples of such complex dynamics.
In turn, functioning of the human brain presents is essence a network activity of the coupled and interrelated
neurons surrounded by glia cells. 
The immense complexity of this subject matter \cite{Pribram} does not exclude, but rather invites thinking
in terms of simple physical modeling approaches, see e.g. in Ref. \cite{Lo1}, 
since even very simple physical models can display
very complex behavior.  The models of critical dynamical phenomena
such as self-organized criticality (SOC) \cite{Sole,SOC,Laszlo} are especially important in this 
respect \cite{Lo2,Lo3}. Physical
modeling  can help to discriminate the physical and biological 
complexity from the complexity of mental processes, the ``form within'' \cite{Pribram},
which is mediated 
but not determined in fine features 
by the background physical processes.  
The recently discovered complexity of the critical brain dynamics
\cite{Beggs,Plentz12,Beggs08} in essence 
does not have anything in common with the complexity of 
mental processes as it is already displayed  by
organotypic networks of neurons formed by cortical slices on a multi-electrode
array \cite{Beggs}. Such physical complexity is in
essence the complexity of crude matter that got self-organized following 
the physical laws. It thus belongs to statistical physics or system biophysics. Physical models 
such as SOC
are   especially important and helpful here. 

The Wilson and Cowan model  \cite{WilsonCowan} presents one
of the well-established models of neuronal network dynamics \cite{Destexhe}. 
It incorporates individual elements in a simplest
possible fashion as two-state stochastic oscillators with one quiescent  state and one excited state, and
random transitions between these two states which are influenced by the mutual coupling among 
the network elements.
The model has been introduced in the deterministic limit of huge many coupled elements in 
 complete neglect 
of the intrinsic mesoscopic noise and became immensely popular with the years \cite{Destexhe}, 
being used e.g. to describe neuronal oscillations in visual cortex within a mean-field
approximation \cite{Sole,Schuster}. 
Recently, the previously neglected 
mesoscopic noise effects were incorporated in this
model for a finite-size network \cite{Benayon,Wallace}. Such a noise has been shown to be 
very important, in particular, for
the occurrence of the critical avalanche dynamics \cite{Benayon} absent in the  
deterministic Wilson-Cowan model and
also for the emergence of oscillatory noisy dynamics \cite{Wallace}. At the first glance,
such a noisy dynamics can look like
a chaotic deterministic one. 
Deterministic chaos influenced by the noise can also be a natural feature
of a higher-dimensional dynamics, beyond the original two-dimensional Wilson-Cowan mean field model. 
Indeed, deterministic chaos has been found in  the brain dynamics
some time ago \cite{Babloyanz,Sole}. However, it cannot be described within the memoryless 
Wilson-Cowan model
because the minimal dimension for chaos is three \cite{Ott}. 

Stochastic mesoscopic noise effects due to a finite
number of elements in finite size systems attract substantial attention over several
decades, especially with respect to chemical reactions on the mesoscale \cite{Gillespie76,
Gillespie00}, being especially pertinent to the physico-chemical 
processes in living cells \cite{Phillips}. In particular, such intrinsic noise can cause and optimize
spontaneous spiking (coherence resonance 
\cite{Pikovsky,LinderReview}) in the excitable clusters of ionic channels in cell membranes, which are
globally coupled through the common membrane potential,
or the response of such systems to periodic external signals (stochastic resonance \cite{Gammaitoni}) 
within a stochastic Hodgkin-Huxley model \cite{Schmid01}. Finite-size networks of 
globally coupled 
bistable stochastic oscillators were also considered without relation to the Wilson-Cowan model 
\cite{Turalska,Pinto}, including
non-Markovian memory effects \cite{Nikitin,Prager06,LinderReview,Prager07,Kouvaris}.

In this paper, we consider a class of 
higher-dimensional generalizations of the
stochastic Wilson-Cowan model aimed to incorporate non-Markovian memory effects in the dynamics
of individual neurons. Such effects are caused by the existence of a 
refractory period or inactivated state
from which the neuron cannot be  excited immediately. First, we discuss a general class of such models.
Then, we apply the simplest two-state non-Markovian model of this 
class, embedded as a three-state Markovian model with one inactivated state, 
to study a critical avalanche dynamics in a balanced network of the excitatory and inhibitory
neurons within a mean-field approximation. 
Here, we restrict ourselves to the simplest example of a fully connected
network with all-to-all coupling
of its elements.
In particular, we derive the power law exponents characterizing the critical self-organized dynamics
of the network from the precise numerical simulations done  with the
dynamic Monte Carlo, or Gillespie algorithm. We also compare these results with similar results obtained
within approximate stochastic
Langevin dynamics, or, equivalently, within a diffusional approximation to the discrete 
state dynamics. Here, we reveal a profound difference.

\section{The model and theory}

\subsection{Stochastic models of single neurons}

\begin{figure}%[!b]
 \centering
 \includegraphics[width=80mm]{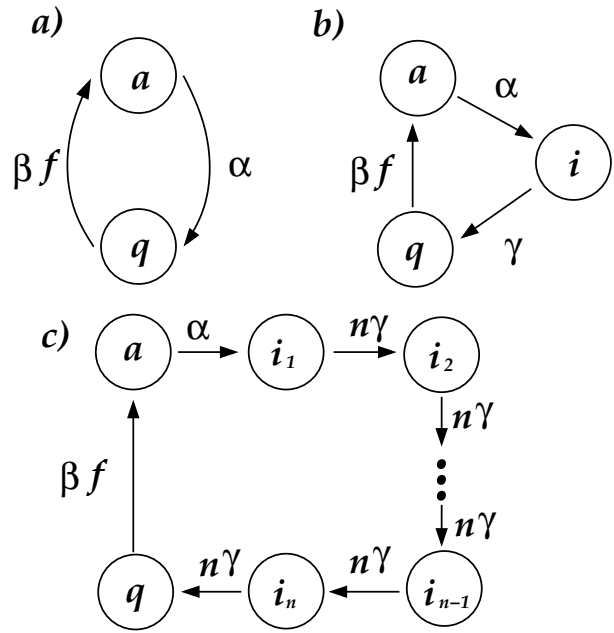}
 \caption{(a) Two-state model of neuron with one quiescent and one excited state.
 (b) Three-state model with one inactivated refractory state. (c) Extension of (b)
 incorporating multi-stage delay.  }
 \label{Fig1} 
\end{figure}

Let us depart from
the Markovian model of a neuron possessing one activated or excited
state ``a'', and a quiescent state ``q'', see Fig. \ref{Fig1}, a. 
The excitation of the neuron occurs with the rate $\beta f$, where $\beta$ is a rate
constant and $f$ is a dimensionless transfer function which depends on the states of the other
neurons, and will be discussed below. Let us assume for a while 
that $f$ is not explicitely time dependent.
From the point of view of the theory of continuous
time random walks (CTRWs) or renewal processes \cite{Cox,Hughes}, 
such a two state neuron can be completely 
characterized by the residence time distributions (RTDs) in its two states, $\psi_a(t)$, and $\psi_q(t)$,
correspondingly 
(assuming that no correlations between the residence time intervals is present -- the renewal
or semi-Markovian assumption). RTDs define completely the trajectory realizations of such a
renewal process.
In the Markovian case, the RTDs are strictly exponential,
$\psi_a(t)=\alpha\exp(-\alpha t)$, and $\psi_q(t)=\nu\exp(-\nu t)$, where
we denoted $\nu=\beta f$. Then, such a trajectory description  corresponds to the
Markovian balance or master equations for the probabilities to populate the states ``a'' and ``q'', $u(t)$ and
$q(t)$, correspondingly. Due to the probability conservation, $u(t)+q(t)=1$,
\begin{eqnarray}\label{markov1}
\dot u(t)=-\alpha u(t) +\beta f[1-u(t)] \;.
\end{eqnarray}
The memory effect due to a delay of a new excitation event 
after the neuron comes into the quiescent state, or 
the existence of some refractory period $\tau_d$, can be captured  within the trajectory description 
by a non-exponential RTD
 $\psi_q(t)$. This transforms  the corresponding master equation into 
a generalized master equation (GME) with memory,
where the term $\beta f q(t)$ is replaced by $\int_{t_0}^t K(t-t')q(t')dt'$, with a memory 
kernel $K(t)$. Here, $t_0$ is the starting time, $t_0=0$, if not a different one is explicitely stated. 
Hence, Eq. (\ref{markov1}) is replaced by 
\begin{eqnarray}\label{non-markov1}
\dot u(t)=-\alpha u(t) + \int_0^t K(t-t') [1-u(t')]dt'\;.
\end{eqnarray}
In the CTRW theory it is well-known how the memory kernel $K(t)$ and the residence time distribution
$\psi_q(t)$ are related \cite{Hughes,Kenkre} (see also Appendix of \cite{Goychuk04}). Namely, 
their Laplace-transforms [denoted as $\tilde F(s)=\int_0^t \exp(-st)F(t)dt$, for
any function $F(t)$] are related as
\begin{eqnarray}\label{Laplace-kernel}
\tilde K(s)=\frac{s\tilde\psi_q(s)}{1-\tilde\psi_q(s)}.
\end{eqnarray}

\begin{figure}
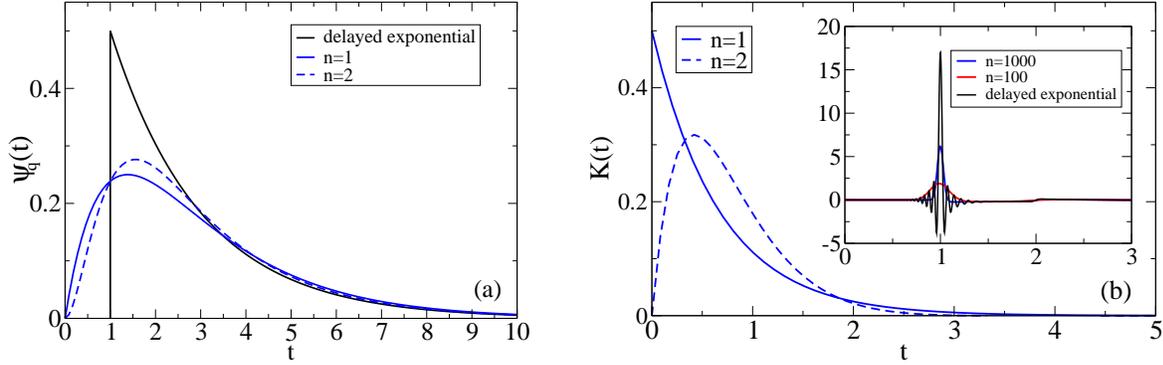

 \centering
 \includegraphics[height=48mm]{Fig2a.eps} \hspace{5mm}
 \includegraphics[height=48mm]{Fig2b.eps}
 \caption{(a) Residence time distribution  in the quiescent state, (b) and the corresponding
 memory kernel for the delayed exponential distribution and distributions corresponding
 to one, $n=1$, and two, $n=2$, inactivated states in Fig. \ref{Fig1}, b, and Fig. \ref{Fig1}, c,
 with $n=2$, correspondingly. 
 Inset in part (b) shows also the cases $n=100$, $n=1000$, and $n\to\infty$ (delayed exponential). 
 Time $t$ is in 
 the units
 of $\tau_d=1/\gamma$, and $\nu=0.5$. Numerical results in the inset were obtained by numerical inversion
 of the corresponding Laplace-transform using the Gaver-Stehfest method with arbitrary precision 
 \cite{Stehfest,Goychuk06} to arrive
 at convergent results.}
 \label{Fig2} 
\end{figure}

In neurosciences, a delayed exponential, or delayed Poissonian model is popular \cite{Koch}.
It is featured by the 
absolute refractory period $\tau_d$, i.e. $\psi_q(t)=0$, for $0\leq t< \tau_d$,
and the exponential distribution, $\psi_q(t)=\nu\exp[-\nu(t-\tau_d)]$, for
$t\ge \tau_d$, see in  Fig. \ref{Fig2}. This model corresponds to 
$\tilde \psi_q(s)=\exp(-\tau_d s)\nu/(s+\nu)$, and the memory kernel 
$\tilde K(s)=\nu s/[(\nu+s)\exp(\tau_d s)-\nu]$. The numerical inverse Laplace transform 
of this memory kernel is depicted in the inset of Fig. \ref{Fig2}, b.  
Notice that it does not correspond to the memory kernel $K(t)=\nu_r\delta (t-\tau_d)$, 
which  would correspond to the master equation with the deterministic delay \cite{Tsimring}
\begin{eqnarray}\label{delay1}
\dot u(t)=-\alpha u(t) +\beta f_r[1-u(t-\tau_d)] \;.
\end{eqnarray}
However, this memory kernel is strongly peaked at $t=\tau_d$, and can  thus be  approximated,
with $\nu_r=\lim_{s\to 0}\tilde K(s)=\nu/(1+\nu\tau_d)$, which is the inverse mean time of
the delayed Poissonian distribution $\psi_q(t)$. In the corresponding Markovian approximation,
Eq. (\ref{markov1}) is restored with a renormalized  transfer function,
\begin{eqnarray}\label{renorm}
f_r=\frac{f}{1+\beta \tau_d f}.
\end{eqnarray}
This is the simplest way to account for the delay effects. Obviously, any delay should 
suppress excitability
within this approximation, because $f_r<f$. However, suppression of the excitability of the inhibitory
neurons may enhance the excitability of the whole network consisting of both excitatory and
inhibitory neurons. Hence, possible effects are generally nontrivial even in this approximation.
Moreover, Eq. (\ref{renorm}) makes it
immediately clear that the delay effects are generally expected to be 
very substantial for $\beta\tau_d\ge 1$.

\subsubsection{The simplest non-Markovian model and its Markovian embedding.} 

It is well-known that in many cases non-Markovian CTRW dynamics can be 
embedded as some Markovian dynamics in a higher-dimensional, possibly infinite dimensional
space \cite{Haus}.
Given a non-trivial form of the memory kernel for the delayed exponential distribution 
of the quiescent times,
we can ask the question: What is the simplest non-Markovian model 
and the corresponding Markovian embedding to account 
for the memory effects?
From the point of view of GME, it is $K(t)=\kappa \exp(-rt)$, i.e. an exponentially decaying
memory kernel. The corresponding memory kernel with $\kappa=\nu\gamma$, and
$r=\nu+\gamma$ corresponds to $\psi_q(t)$ , which is the time convolution 
of two exponential distributions, $\psi_q^{(0)}(t)=\nu\exp(-\nu t)$, and 
$\psi_i(t)=\gamma \exp(-\gamma t)$. It corresponds to a compound state ``$\mathbf{q}=q\cup i$''
in Fig. \ref{Fig1}, b.
Indeed, the Laplace transform of the corresponding compound distribution is just the product
of the Laplace-transforms of two exponential distributions, i.e., 
$\tilde \psi_q(s)=\nu\gamma/[(s+\nu)(s+\gamma)]$. By Eq. (\ref{Laplace-kernel}) it corresponds precisely
to the stated exponential memory kernel. The corresponding 
$\psi_q(t)=\nu\gamma[\exp(-\nu t)-\exp(-\gamma t)]/(\nu-\gamma)$ has 
a maximum at the most probable time interval $t_{\rm max}=\ln(\nu/\gamma)/(\nu-\gamma)$,
see Fig. \ref{Fig2}, a, reflecting the most probable stochastic time delay.
This simplest non-Markovian model with memory allows, however,
for a very simple Markovian embedding by introduction of an intermediated refractory
state ``i'' shown in Fig. \ref{Fig1}, b, with the population probability $x(t)$ and
the exponential RTD given above. It has the mean refractory time 
$\tau_d:=\langle \tau \rangle =\int_0^\infty \psi_i(\tau)d\tau=1/\gamma$, and the
relative standard deviation, or the coefficient of variation, $C_V:=\sqrt{\langle  \tau^2 \rangle-
\langle \tau \rangle^2}/ \langle \tau \rangle=1$. Using the
conservation law, $u+q+x=1$, the corresponding master equations can be written either
as
\begin{eqnarray}\label{3state1}
\dot u & = &-\alpha u + \nu (1-u-x),\nonumber \\
\dot x & = & \alpha u-\gamma x ,
\end{eqnarray}
or as
\begin{subequations}
\begin{eqnarray}\label{3state2a}
\dot u & = &-\alpha u + \nu q, \\
\dot q & = & -(\gamma+\nu) q+\gamma(1-u)\;.\label{3state2b}
\end{eqnarray}
\end{subequations}
From (\ref{3state2b}) follows 
\begin{eqnarray}
q(t)=e^{-(\gamma+\nu)t}q(0)+\gamma\int_0^t e^{-(\gamma+\nu)(t-t')}[1-u(t')]dt'\;.
\end{eqnarray}
After substitution of this equation into (\ref{3state2a}) one obtains indeed Eq. (\ref{non-markov1})
with the discussed exponential memory kernel provided that $q(0)=0$. The latter condition
is natural because every sojourn in the compound quiescent state ``$\mathbf{q}$''
starts from the substate ``i'' (resetting memory of this neuron to zero), and $q(t)+x(t)$ is 
the probability of the compound
quiescent state within the two-state non-Markovian reduction 
of the three-state Markovian problem. Here, one can also see  the origin
of a profound problem with the description  of the whole network dynamics of interacting 
non-Markovian renewal elements  as a hyper-dimensional renewal process. Obviously,
the behavior of the whole network cannot be considered as a renewal process, because
after each and every de-excitation event only one element is reset. Then it starts with
zero memory, while all others  keep their memory until they are reset. 
Hence, any Gillespie type simulation of the whole network 
of interacting non-Markovian elements must account for the ``age'' of each network element separately.
Markovian embedding allows to circumvent this problem and dramatically accelerate 
simulations within the mean-field approximation, see below. 

The considered three-state Markovian
cyclic model presents one of the fundamental kinetic models in biophysics. 
It provides, in particular, a paradigm for non-equilibrium steady state cycling. For example,
the cyclic kinetics
of an enzyme $E$, which binds a substrate molecule $S$, converts it to a product $P$, and releases it
afterwards
can be represented as a three-state cycle, $E\to ES \to EP \to E$. 
This model was used e.g. in Ref. \cite{Prager06} for an excitable
unit. Furthermore, three-state non-Markovian models
can be used with a non-exponential distribution $\psi_i(t)$. 
For example, if to use the deterministically
delayed $\psi_i(t)=\delta(t-\tau_d)$, and exponential $\psi_q(t)$ within the three-state cyclic model, 
then one
obtains the  delayed exponential distribution within the two-state reduced model,
which was discussed above. In addition, $\psi_a(t)$ can also be non-exponential.
For $\psi_a(t)=\delta(t-1/\alpha)$ in the excited state of the three-state non-Markovian model, 
one obtains  the model used in Refs. \cite{Nikitin,Prager06}.

\subsubsection{Markovian embedding with many substates.} 

One can also introduce many delayed substates as shown in Fig. \ref{Fig1}, c. 
Within the three-state non-Markovian model this can be considered
as having one delayed state ``i'' characterized by the special
Erlangian distribution \cite{Cox}, 
$\psi_i(t)=n\gamma(n\gamma t)^{n-1}\exp(-n\gamma t)/(n-1)!$, with the Laplace-transform 
$\tilde\psi_i(s)=1/(1+s/(n\gamma))^n$ reflecting the corresponding multiple convolution. 
Such a non-Markovian three-state model has been considered in \cite{Prager07}, and
a non-Markovian two-state model with the Erlangian distribution of the quiescent times 
has been studied in \cite{Kouvaris}.
The compound quiescent state corresponding to the model in \cite{Prager07} 
is characterized by $\tilde\psi_q(s)=\nu/[(1+s/(n\gamma))^n(\nu+s)]$.
The mean delay time is the same $\tau_d=1/\gamma$ for any $n$, and the coefficient 
of variation becomes ever smaller with increasing $n$, $C_V=1/\sqrt{n}$, i.e. the distribution
of the refractory times becomes ever more sharpened. 
The Laplace-transformed memory kernel is $\tilde K(s)=\nu s/[(1+s/(n\gamma))^n(\nu+s)-\nu]$.
Some corresponding $\psi_q(t)$ and $K(t)$ are shown in Fig. \ref{Fig2}. Already for $n=2$,
the memory kernel starts to show a peaked structure.
Notice that in the limit $n\to\infty$, the above delayed exponential (or Poissonian) model
immediately follows with $\tau_d=1/\gamma$. For any $n$, the inverse mean time
in the quiescent state is given by $\beta f_r$ with $f_r$ in (\ref{renorm}). 
Increasing $n$ yields an ever better approximation
for the delayed Poissonian model. However, it can be considered as a useful model in itself. 
The corresponding Markovian embedding master equation reads (with $q$ 
excluded by the probability conservation law):
\begin{eqnarray}\label{3state_n}
\dot u & = &-\alpha u + \beta f \left(1-u-\sum_{i=1}^n x_i \right ),\nonumber \\
\dot x_1 & = & \alpha u-n\gamma x_1,\nonumber \\
\dot x_j & = & n\gamma(x_{j-1}-x_j), j=2,..,n\;,
\end{eqnarray}
with, $x_j(0)=0$, for $j=2,...,n$, initially. With a different initial condition, the
corresponding GME obtained upon projection of the multi-dimensional dynamics onto the subspace
of $u$ and $q$ variables will contain a dependence on this initial condition in the subspace of
hidden Markovian variables. On the level of non-Markovian dynamics this can be accounted
for by a different choice of the residence time distribution $\psi_q^{(0)}(t)$ for 
the first sojourn in the quiescent state. It depends on how long this state has been populated
before the dynamics started \cite{Cox}. The GME (\ref{non-markov1}), 
(\ref{Laplace-kernel}) corresponds to the particular choice, $\psi_q^{(0)}(t)=\psi_q(t)$.

We mention in passing also that it is straightforward to consider a power-law distributed
delay, both within a semi-Markovian model and within an approximate finite-dimensional
Markovian embedding. Also a stochastic model for bursting neurons can be introduced
immediately. We shall not, however, consider these possibilities in the present work.

\subsection{Network of neurons within the mean field dynamics}

Following Wilson and Cowan, we consider a network of $N_e$  excitatory and 
$N_i$ inhibitory neurons, with the probabilities of neurons to be in their excited
states $u_i(t)$ and $v_i(t)$, correspondingly. The neuron $k$ can 
influence the neuron $l$ and possibly itself ($k=l$), by excitation, or inhibition
with the coupling
constants, $w^{lk}_{ee}>0$, $w^{lk}_{ie}>0$ for the excitatory neuron $k$, and 
$-w^{lk}_{ei}<0$, $-w^{lk}_{ii}<0$, for the inhibitory neuron $k$.
The absolute value of the coupling constant  reflects the synaptic strength. 
Each excitatory 
neuron $l$  thus obtains an  averaged input $s_l=(1/N^{l}_e)\sum_k w^{lk}_{ee}u_k-
(1/N^{l}_i)\sum_p  w^{lp}_{ei}v_p+h_e^l$,
and the inhibitory neuron $p$ receives the input $s_p=(1/N^{p}_e)\sum_l w^{pl}_{ie}u_l-(1/N^{p}_i)
\sum_k  w^{pk}_{ii}v_k+h_i^p$, where $N^{l}_e$ and $N^{l}_i$, etc. is the number of inputs
which the $l-$th neuron obtains from the excitatory and inhibitory neurons, correspondingly.
The constants $h_e^l$ and $h_i^p$ serve to fix
the spontaneous spiking rates, $\beta_l f(h_e^l)$, $\beta_p f(h_i^p)$, in the absence of coupling,
 $w^{lk}_{ee}\to 0,w^{lk}_{ii}\to 0,w^{lk}_{ei}\to 0,w^{lk}_{ie}\to 0$.
Coupling can either enhance, or suppress these rates. 
%Synaptic input can move effectively the excitation threshold  either up (inhibition), or
%down (excitation). 
Phenomenologically, this is accounted for by the transfer function $f(s)$,
which we assume to be the same for all neurons.
Some common biophysically motivated popular choices of the transfer function $f(s)$ are
\begin{eqnarray}\label{t1}
f(s)=\tanh(s)\theta(s), 
\end{eqnarray}
where $\theta(s)$ is the Heaviside step function, and 
\begin{eqnarray}\label{t2}
f(s)=1/[1+\exp(-s)]\;.
\end{eqnarray}
Both are bounded as $0\leq f \leq 1$.  Evidently, this is a very
rich model even for the simplest two-state model of neurons, in the absence of memory
effects. The simplest further approximation to describe the collective dynamics of neurons
is to invoke the mean field approximation \cite{Schuster}. It is equivalent
to assuming that all the coupling constants like $w^{lk}_{ee}$, etc., thresholds $h_e^l$,
etc., and rates 
$\beta_l$, $\alpha_l$, are equal within a subpopulation,  $w^{lk}_{ee}=w_{ee}$, 
$h_e^l=h_e$, $\beta_{e}^l=\beta_e$, or $\beta_i^l=\beta_i$, etc. Furthermore,
one can introduce the occupation numbers  of the excited neurons in each population, 
$u(t)=(1/N_e)\sum_{i=1}^{N_e}u_i(t)$, and $v(t)=(1/N_i)\sum_{i=1}^{N_i}v_i(t)$, 
and consider the dynamics of these variables. They present the fractions of neurons
which are excited.

We restrict our treatment
in the rest of this paper to the simplest two state non-Markovian model within the three state Markovian embedding
and introduce the occupation numbers of neurons, $x(t)=(1/N_e)\sum_{i=1}^{N_e}x_i(t)$, 
and $y(t)=(1/N_i)\sum_{i=1}^{N_i}y_i(t)$, in the corresponding delayed states. Then, in the deterministic
limit $N_e,N_i\to\infty$, one obtains a 4-dimensional nonlinear dynamics,
\begin{eqnarray}\label{4dim}
\dot u & = &-\alpha_e u + \beta_e f(w_{ee} u-w_{ei} v+h_e) (1-u-x),\nonumber \\
\dot x & = & \alpha_e u-\gamma_e x ,\nonumber \\
\dot v & = &-\alpha_i v  + \beta_i f(w_{ie} u-w_{ii} v+h_i) (1-v-y),\nonumber \\
\dot y & = & \alpha_i v-\gamma_i y\;.
\end{eqnarray}
Notice that unlike the original two-dimensional mean-field Wilson-Cowan dynamics in the deterministic
limit, the considered 4-dimensional dynamics can in principle be chaotic, for some parameters 
(which remains an open question). Dynamical chaos might emerge already
when only one sort of neurons, e.g.
inhibitory neurons, exhibits delayed dynamics, since the minimal dimension for nonlinear chaotic
dynamics is three. Then, in the macroscopic deterministic limit,
\begin{eqnarray}\label{3dim}
\dot u & = &-\alpha_e u + \beta_e f(w_{ee} u-w_{ei} v+h_e) (1-u),\nonumber \\
\dot v & = &-\alpha_i v  + \beta_i f(w_{ie} u-w_{ii} v+h_i) (1-v-y),\nonumber \\
\dot y & = & \alpha_i v-\gamma_i y\;.
\end{eqnarray}
However, we shall not
investigate the possibility of a deterministic chaos emerging due to a delay within the minimal
extensions of the Wilson-Cowan model
in the present work, but rather focus on the mesoscopic intrinsic
noise effects caused by finite $N_e$ and $N_i$. Then, the occupational numbers
are random variables (at any fixed time $t$). 

\subsubsection{Langevin dynamics.}
For a very large number of neurons, one
can account for the mesoscopic noise effect within the Langevin dynamics,
or the diffusional approximation of the discrete state birth-and-death process describing the evolution
of the network.  This procedure
is standard, by analogy with the stochastic theory of chemical reactions \cite{Gillespie00}.
Since we have only direct ``reactions'' like $q\rightarrow a$, $a\rightarrow i$,
$i\rightarrow q$, for two type of neurons, one must introduce six variables and six
 independent zero-mean white Gaussian
noise sources $\xi_i(t)$, $\langle \xi_i(t)\xi_j(t')\rangle=\delta_{ij}\delta(t-t')$. 
Stochastic dynamics is, however, effectively 4-dimensional because of two probability 
conservation laws, which allow
to exclude two variables out of six:
\begin{eqnarray}\label{4dimLangevin}
\dot u & = &-\alpha_e u + \beta_e f(w_{ee} u-w_{ei} v+h_e) (1-u-x)\nonumber \\&& -
\frac{1}{\sqrt{N_e}}\sqrt{\alpha_e u}\xi_1(t)+\frac{1}{\sqrt{N_e}}
\sqrt{\beta_e f(w_{ee} u-w_{ei} v+h_e) (1-u-x)}\xi_2(t),\nonumber \\
\dot x & = & \alpha_e u-\gamma_e x \nonumber +
\frac{1}{\sqrt{N_e}}\sqrt{\alpha_e u}\xi_1(t)-\frac{1}{\sqrt{N_e}}\sqrt{\gamma_e x}\xi_3(t),\\
\dot v & = &-\alpha_i v  + \beta_i f(w_{ie} u-w_{ii} v+h_i) (1-v-y)\nonumber \\ && -
\frac{1}{\sqrt{N_i}}\sqrt{\alpha_i v}\xi_4(t)+\frac{1}{\sqrt{N_i}}
\sqrt{\beta_i f(w_{ie} u-w_{ii} v+h_i) (1-v-y)}\xi_5(t),\nonumber \\
\dot y & = & \alpha_i v-\gamma_i y + 
\frac{1}{\sqrt{N_i}}\sqrt{\alpha_i v}\xi_4(t)-\frac{1}{\sqrt{N_i}}\sqrt{\gamma_i y}\xi_6(t)\;.
\end{eqnarray}
In the limit $N_e,N_i\to\infty$, the deterministic description in Eq. (\ref{4dim})
is restored. The noise is multiplicative and the Langevin
equations must be Ito-interpreted, as it is always the case if the Langevin
dynamics results from the standard  diffusional approximation to a birth-and-death process, 
or chemical master equation \cite{Gillespie00}.
Notice that such a Langevin stochastic description can become problematic, 
if any of the variables $u,v,x,y$ becomes temporally zero or one.
Even if some of  the noise terms do vanish at the boundaries, where there corresponding rates
vanish,  the others do not, when a particular boundary is hit. Hence, the occupational numbers
can in principle become temporally negative, 
or larger than one. This unphysical feature is produced by the
 standard diffusional approximation. 
However, this problem can be fixed  in the numerical simulations by introduction of
the corresponding reflecting boundaries and taking sufficiently small integration time
steps, as done e.g. in Ref. \cite{Schmid01,Goychuk14} for stochastic Hodgkin-Huxley equations.

\subsection{Exact stochastic simulations.}

\begin{table} % add [H] placement to break table across pages
\begin{center}
\caption{Transitions and rates}\label{Table1}\vspace{0.5cm}

%\begin{ruledtabular}
\begin{tabular}{|p{0.3cm}|p{7.3cm}|p{6.5cm}|}
\hline
$i$ & Transition & Rate $r_i$\\
\hline
1 & $(N,M,N_1,M_1)\rightarrow (N-1,M,N_1+1,M_1)$ & $r_1=N\alpha_e$ \\
\hline
2 & $(N,M,N_1,M_1)\rightarrow (N,M-1,N_1,M_1+1)$ & $r_2=M\alpha_i$ \\
 \hline
3 & $(N,M,N_1,M_1)\rightarrow (N+1,M,N_1,M_1)$ & $r_3=(N_e-N-N_1)\beta_e f(w_{ee} N/N_e-w_{ei}M/N_i+h_e)$ \\
\hline
4 & $(N,M,N_1,M_1)
\rightarrow (N,M+1,N_1,M_1)$ & $r_4=(N_i-M-M_1)\beta_i f(w_{ie} N/N_e-w_{ii}M/N_i+h_i)$ \\
\hline
5 &  
$(N,M,N_1,M_1)\rightarrow (N,M,N_1-1,M_1)$ & 
$r_5=\gamma_e N_1$ \\
\hline 
6 & 
$(N,M,N_1,M_1)\rightarrow (N,M,N_1,M_1-1)$ & 
$r_6=\gamma_i M_1$ \\
\hline
\end{tabular}
\end{center}
%\end{ruledtabular}
\end{table}

Within the mean-field approximation of Markovian dynamics, it suffices to 
count the numbers of neurons
in the corresponding  activated, $N$ and $M$, and refractory, $N_1$ and $M_1$ states. 
Then, we are dealing with a random walk on a 4-dimensional lattice $(N,M,N_1,M_1)$ 
with the discrete variables 
$N$, and $N_1$ taking values in the range from zero to $N_e$, and the variables 
$M$ and $M_1$ in the range from zero to $N_i$, so that also $0\le N+N_1\le N_e$ 
and $0\le M+M_1\le N_i$. From the site $(N,M,N_1,M_1)$  six different
transitions  are possible. They are enlisted in 
 Table \ref{Table1} with the corresponding transition
rates. The master equation governing this birth-and-death process can
be readily written. However, it is bulky and not very insightful. For this reason, it 
is not presented here.
The corresponding
stochastic process can be easily simulated with the dynamical Monte Carlo
 or Gillespie algorithm \cite{Gillespie76},
which is exact. Namely, on each step one draws two random numbers. The first one, 
$\tau$, is drawn from the exponential
distribution characterized by the total rate $r_\Sigma=\sum_{i=1}^6 r_i$. 
It gives a random time interval at which 
the network state is updated.  Given a uniformly distributed random variable $\zeta_1$, 
$0 \le \zeta_1 \le 1$, $\tau=(1/r_\Sigma)\ln (1/\zeta_1)$. Then, one of the  transitions 
in  Table~ \ref{Table1} is chosen in
accordance with its probability $p_i=r_i/r_\Sigma$.
For this, one generates 
a uniformly distributed random variable  $\zeta_2$ bounded as
$0 \le \zeta_2 \le 1$. If $0<\zeta_2 <p_1$, then the first transition is chosen.
If $p_1\leq \zeta_2 <p_1+p_2$, then the second transition is chosen, etc., i.e. in accordance
with the length of the corresponding interval $p_i$, $\sum_{i=1}^6 p_i=1$. 

Notice that an attempt to generalize this scheme towards a non-Markovian renewal 
walk on a 4-dimensional lattice to account for the memory in the inactivated
state is logically inconsistent because in such a case accomplishing
 each step would mean reset, or renewal of \textit{all}
neurons, and not the only one which actually makes  transition. 
However, each non-Markovian element 
has its individual memory. 
In a direct simulation of the network of
non-Markovian elements one must therefore consider them individually, even within the
mean-field approximation. Then, 
one has to consider CTRW on a hyper-dimensional lattice of huge dimensionality, which will
dramatically  slow down
simulations imposing computational restrictions on the maximal size of the network.
Of course, beyond the mean field approximation one must also simulate each element separately.
Here, a direct semi-Markovian approach can be preferred. In this work, we restrict ourselves to 
the mean-field dynamics within the Markovian embedding framework, which allows for
exact simulations of very large networks within a reasonable computational time.

\section{Results and Discussion}

\subsection{Oscillatory dynamics of neuronal network}

\begin{figure}
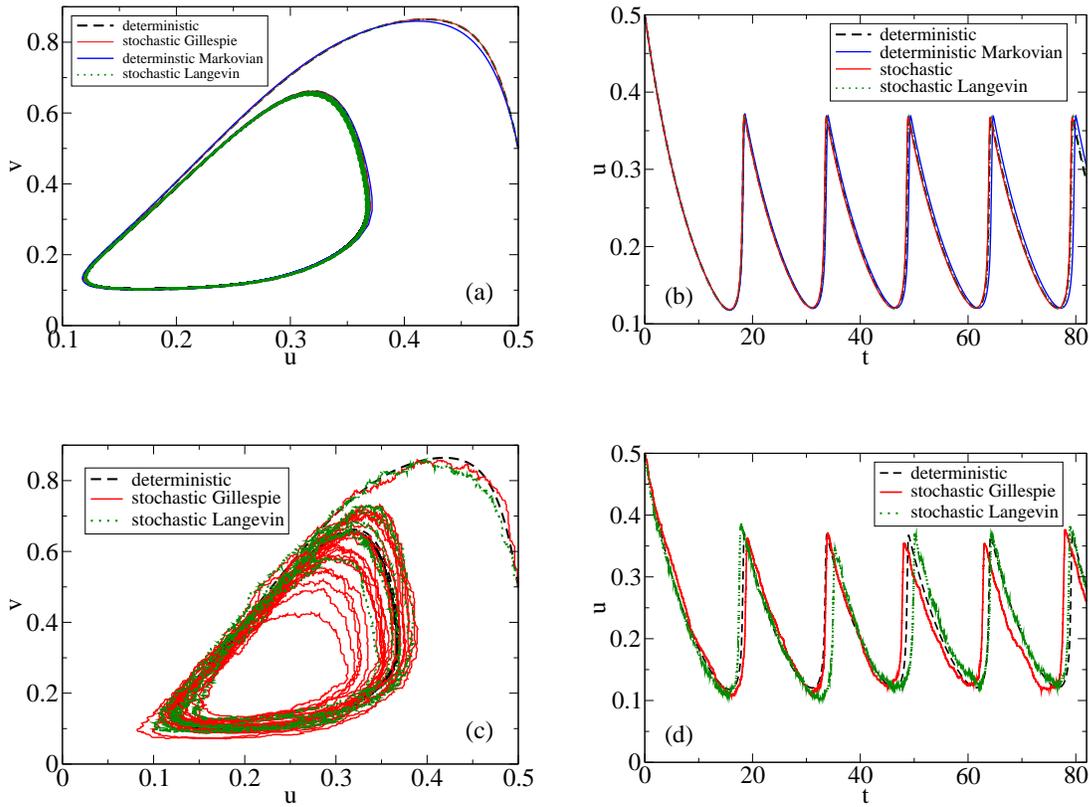

 \centering
 \vspace{0.5cm}
 \includegraphics[height=48mm]{Fig3a.eps} \hspace{5mm}
 \includegraphics[height=48mm]{Fig3b.eps}\\ \vspace{1cm}
 
 \includegraphics[height=48mm]{Fig3c.eps} \hspace{5mm}
 \includegraphics[height=48mm]{Fig3d.eps}
 \caption{(a) Limiting cycle in the $u-v$ plane for deterministic dynamics
 and for stochastic dynamics, $u=N/N_e$, $v=M/N_i$, with $N_e=N_i=10^6$, (b) Time-dependence of the
 $u$ variable for (a). (c) Limiting cycle in the $u-v$ plane for deterministic dynamics
 and for stochastic dynamics  with $N_e=N_i=10^3$, and (d) the corresponding time-dependence
 of the $u$ variable. Langevin simulations are done with the stochastic Euler algorithm
 using time step $\delta t=0.05$ in (a) and (b), and $\delta t=10^{-5}$ in (c) and (d).}
 \label{Fig3} 
\end{figure}

\begin{figure}
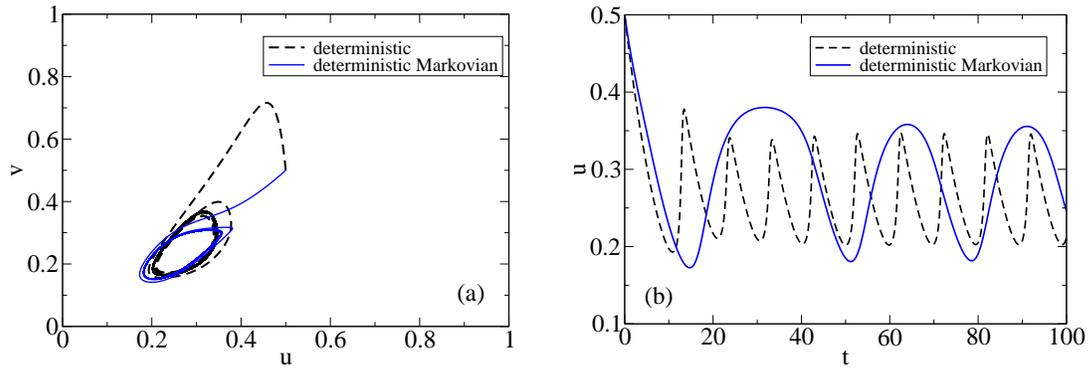

 \centering
 \vspace{0.5cm}
 \includegraphics[height=48mm]{Fig4a.eps} \hspace{5mm}
 \includegraphics[height=48mm]{Fig4b.eps}\\ \vspace{0.5cm}
 \caption{Deterministic two-state dynamics with memory  and its Markovian 
 approximation for $\gamma_e=\gamma_i=0.1$. Other parameters are 
 the same as in Fig. \ref{Fig3}. The deviation in (b) indicates that
 the Markovian approximation fails completely.}
 \label{Fig4} 
\end{figure}

First, we test our stochastic simulations done with  XPPAUT \cite{Bard} against nonlinear
deterministic dynamics for a very large network size with
$N_e=10^6$ and $N_i=10^6$. For this, departing from the parameter set in 
Ref. \cite{Wallace} (the case of without delay) 
we use a set of parameters, where an oscillatory
dynamics emerges: $\alpha_e=0.1$, $\alpha_i=0.2$, $\beta_e=1$, $\beta_i=2$, 
$\gamma_e=10$, $\gamma_i=10$, $w_{ee}=32$, $w_{ei}=32$, $w_{ie}=28$, $w_{ii}=2$,
$h_e=-3.8$, $h_i=-9$, and the transfer function in Eq. (\ref{t2}). 
Time is in milliseconds and the  rate constants are in inverse milliseconds. The difference
is barely detectable in Fig. \ref{Fig3}, a, b, where
we present the results of stochastic simulations done both with the exact Gillespie
algorithm and within the approximate Langevin dynamics.
However, stochastic effects become
immediately seen in Fig. \ref{Fig3}, c, d, where we reduced the number of neurons to
$N_e=10^3$ and $N_i=10^3$. We also compare in Fig. \ref{Fig3}, a, b, the results
for the considered dynamics and its two-variable Markovian 
approximation given by the standard Wilson-Cowan model in 
which, however, the transfer functions are renormalized in accordance with Eq. (\ref{renorm}), where
the parameter $\beta\tau_d$ is replaced by $\beta_e/\gamma_e=0.1$ and $\beta_i/\gamma_i=0.2$, correspondingly.
The deviations are visible, but small. 
However, the differences become very
pronounced for small 
$\gamma_e=\gamma_i=0.1$ corresponding to the mean refractory period $\tau_d=1/\gamma_{e,i}$ 
of 10 msec.
Then, the Markovian approximation fails completely, see in Fig. \ref{Fig4}, especially in part (b),
revealing that neither the form of the oscillations, not their period are reproduced even approximately. 
Especially remarkable is that contrary to intuition the increase of 
the refractory period of a single neuron
does not increase the period of oscillations, as the Markovian approximation predicts, but
rather makes it smaller -- the tendency is opposite! 
Therefore, non-Markovian memory effects generally do matter and one should take such effects
 seriously into account.
With a small further decrease of
 $\gamma_e,\gamma_i$ to $\gamma_e=\gamma_i\approx 0.08873$ with $\tau_d\approx 11.270$ the oscillations
 are terminated by a supercritical Hopf bifurcation (not shown). Interestingly, Markovian approximation 
 also predicts such a termination,
 but at a slightly larger critical value $\gamma_e=\gamma_i\approx 0.0987$ with
 critical $\tau_d\approx 10.132$. This makes clear that the phase transitions between the quiescent 
 network and the network undergoing synchronized oscillations  are possible with respect
 to the length of the refractory period used as a control parameter.

\subsection{Noise-induced critical avalanche dynamics}

\begin{figure}
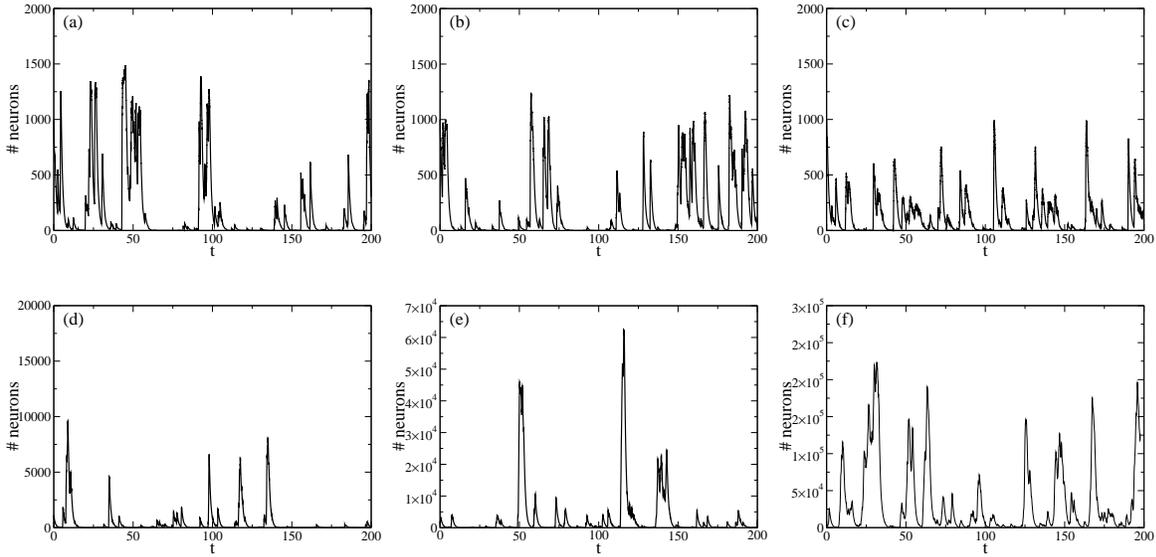

 \centering
 \vspace{0.5cm}
 \includegraphics[width=5cm]{Fig5a.eps} 
 \includegraphics[width=5cm]{Fig5b.eps}
 \includegraphics[width=5cm]{Fig5c.eps}\\ \vspace{0.5cm}
 \includegraphics[width=5cm]{Fig5d.eps} 
 \includegraphics[width=5cm]{Fig5e.eps}
 \includegraphics[width=5cm]{Fig5f.eps}\\ \vspace{0.5cm}
  \hspace{5mm}
 \caption{Avalanches for $N_e=N_i=10^3$ and (a) $\gamma_e=\gamma_i=10$, (b) $\gamma_e=\gamma_i=1$,
 (c) $\gamma_e=\gamma_i=0.1$.
 In (d), (e), and (f), $\gamma_e=\gamma_i=10$, however, the network size is increased to
 (d) $N_e=N_i=10^4$, (e) $N_e=N_i=10^5$, and (f) $N_e=N_i=10^6$. 
 For the fixed $N_e=N_i=10^3$, the maximal
 $L(t)$ in (a) is $1513$, i.e. about 76\% of maximally possible. With the increase of refractory
 time it diminishes to $1322$ (about  66\%) in (b), and to $1169$ (about 58.5\%) in (c).
 For a fixed refractory time, but with the increase of the network size 
 the maximal number of active neurons is $12791$ ($\sim$64\%) in (d), $61923$ ($\sim$31\%) in (e),
 and $223234$ ($\sim$11\%) in (f). With the increase of network size, the relative size of avalanches
 decreases.  Notice also that the minimal number of active neurons is
 $L_{ \rm min}=3$ in (e), and  $L_{\rm min}=112$ in (f). This must be taken into account
 when one defines avalanches in large networks. Otherwise, one can come to incorrect conclusion
 that the avalanches cease to exist, which is manifestly refuted in (f) for a very large number 
 of neurons which seems to be macroscopically large, and nevertheless fluctuations are still very
 important, even though they do vanish in the strict limit $N_e,N_i\to\infty$. 
 Experimentally, one also defines the start and end of an avalanche by crossing
 a threshold of basal activity upwards, and downwards, correspondingly. Simulations
 are done with the Gillespie algorithm.}
 \label{Fig5} 
\end{figure}

In the remainder, we  investigate the influence of memory effects on the avalanche dynamics.
 As it has been shown in \cite{Benayon}, in  order to
have avalanche dynamics the excitatory and inhibitory processes should
be nearly balanced, and the network should
have a so-called feedforward structure. Then, one can achieve a sort of self-organized critical (SOC) state 
\cite{SOC,Laszlo}
sustained due to intrinsic mesoscopic fluctuations. Very different from other SOC models,
here fluctuations play a major role and in the deterministic limit avalanches disappear,
i.e. they are of mesoscopic nature.
The  nullclines of 2d deterministic dynamics in the absence of memory effects,
$\dot u=0$ and $\dot v=0$, should cross at a very small angle
in the $u-v$ plane, so that fluctuations can produce large amplitude outbursts of the
$u$ and $v$ variables moving synchronously but randomly, i.e. the subpopulations of excitatory
and inhibitory neurons are synchronized exhibiting stochastic dynamics at the same time \cite{Benayon}.
In the same spirit,
we choose $\alpha_e=\alpha_i=1$, $\beta_e=\beta_i=5$, 
$w_{ee}=w_{ie}=w_e=30$, $w_{ei}=w_{ii}=w_i=29.9$, so that $w_e-w_i\ll w_e+w_i$, and overall excitation
 slightly dominates over inhibition. Furthermore, we choose
$h_e=h_i=0.001$, and the transfer function in Eq. (\ref{t1}), as in Ref. \cite{Benayon}. 
The rates 
$\gamma_e$, $\gamma_i$ and the number of neurons  were varied.  Large 
$\gamma_e=\gamma_i=10$ corresponds to a small refractory time of $0.1$ msec, 
whereas $\gamma_e=\gamma_i=0.1$ corresponds
to a profound delay with $\tau_d=10$ msec, so that the individual spiking rate
of neurons cannot exceed 100 Hz  being limited by the refractory period. Typical avalanche
dynamics is shown in Fig. \ref{Fig5} for $L(t)=N(t)+M(t)$ with $N_e=N_i=10^3$, and (a) 
$\gamma_e=\gamma_i=10$,  (b) $\gamma_e=\gamma_i=1$, (c) $\gamma_e=\gamma_i=0.1$.
Furthermore, in Fig. \ref{Fig5}, d, e, f, we show the influence of an increasing number of neurons
on the avalanche dynamics. The following tendencies are clear. First, the increase of refractory
period reduces the maximal amount of neurons involved in spiking, from about 76\% in Fig. \ref{Fig5}, a
to  58.5\% in Fig. \ref{Fig5}, c. Such a tendency 
is already expected  from the renormalization of the transfer function
in the Markovian approximation, cf . Eq. (\ref{renorm}). However, this tendency is in fact
much weaker since $\beta\tau_d=50$ in the part (c),
and the renormalization argumentation would predict almost complete suppression of avalanches for
such a delay. Even more
astonishing is that avalanches still did not vanish even for a very large
$N_i=N_e=10^6$, see Fig. \ref{Fig5}, f. This is very different from the oscillatory dynamics
of a network of the same size, cf. Fig. \ref{Fig3}, a, and b, which is practically deterministic.
Of course, with  increasing network size, the relative amplitude of  avalanches becomes ever smaller,
and there also emerges a minimal number of neurons  excited, i.e. the 
network activity never goes
down to zero. However, this is also so in the real neuronal dynamics. 
Such a dominance of mesoscopic fluctuations in a system of millions elements with  a special
(feedforward) structure of coupling is really astonishing.
This is a feature of some critical state, as we know from statistical physics. 

\begin{figure}
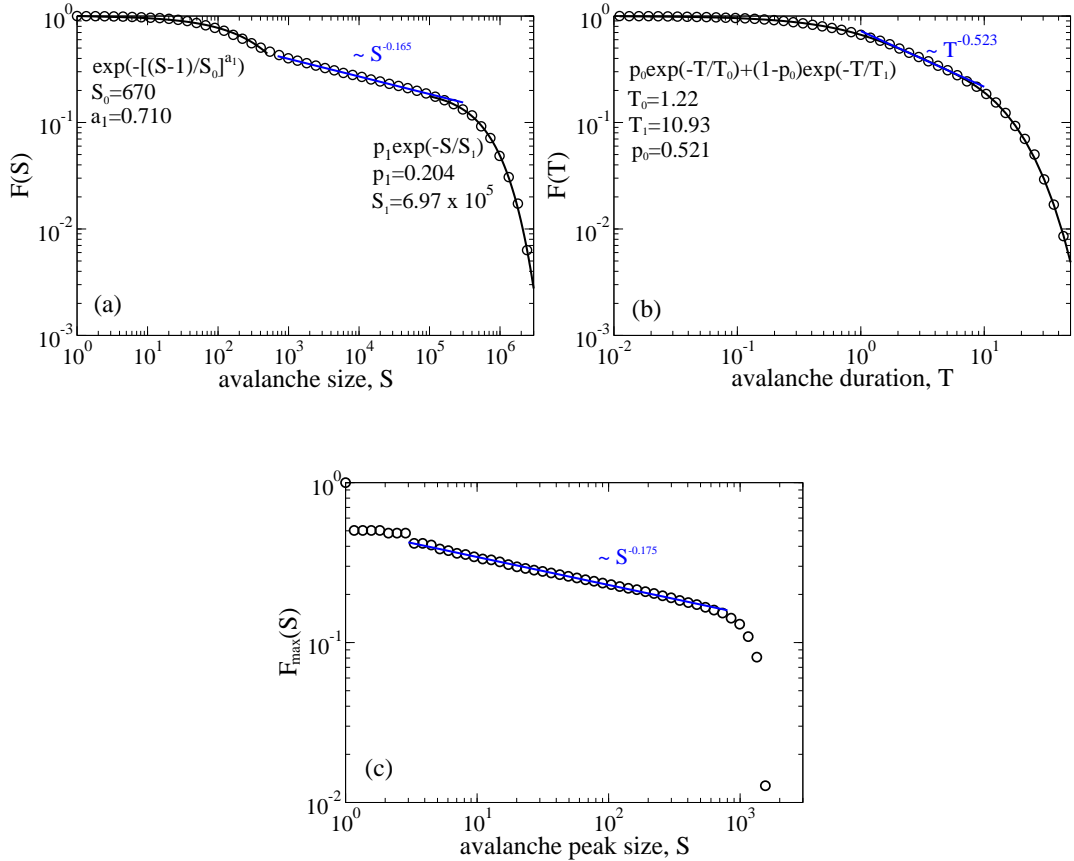

 \centering
 \vspace{0.5cm}
 \includegraphics[width=7cm]{Fig6a.eps} 
 \includegraphics[width=7cm]{Fig6b.eps}
 \vfill \vspace{1cm}
 \includegraphics[width=7cm]{Fig6c.eps}
 \caption{Survival probabilities for the avalanche size (a), duration (b), and peak (c)
 distributions obtained from the Gillespie simulations. The case without delay. Circle symbols correspond to numerical
 results and lines to the corresponding fits with parameters displayed on the plot. Blue line
 corresponds to the power law fit. $N_e=N_i=10^3$. Other parameter are given in the text.}
 \label{Fig6} 
\end{figure}

To statistically characterize  the avalanche size distribution and their duration we proceed
in accordance with the procedure outlined in Ref. \cite{Benayon}. It reflects, in part, also
the experimental procedure \cite{Plentz12}. Namely, we first discretize the time series with a time bin of 
the size $\langle \Delta t \rangle$, which corresponds to the averaged interspike time 
distance in a particular simulation. Then, an avalanche is defined by its start, when the spiking
activity crosses some threshold level $L_{\rm thr}$,  and its end, when the network activity
drops to ($L_{\rm thr}=0$) or below $L_{\rm thr}>0$
after some time, which defines the avalanche duration.
The size is defined as the sum of the number of neurons active in each time bin during the
avalanche.  It is also defined experimentally in such a way.
In essence, the size  $S$ of an avalanche is the integral of the network activity in Fig. \ref{Fig5} 
over the time during each avalanche period divided by the time bin width.  Of course, 
as also in experiments
the critical exponents discussed below depend both on the time bin width and  on the basal level
of neuronal activity $L_{\rm thr}$. However, this dependence is weak for
a truly critical dynamics.
By doing statistical analysis, we first find  the survival probability $F(S)$, or, equivalently,
the cumulative probability
$1-F(S)$ from the numerical data. Then, the distribution density follows as $p(S)=-d F(S)/dS$.

Let us start from the case without any time delay, $\tau_d=0$.
The survival probability for the avalanche size distribution $F(S)$ is shown in Fig. \ref{Fig6},a,
for the time bin $\langle \Delta t \rangle=0.00643597$
and $L_{\rm thr}=0$.
It shows three characteristic features: (1) an initial Weibull distribution, 
$F(S)=\exp(-[(S-1)^{a_1}/S_{0}^{a_1}])$, with $a_1\approx 0.71$ and $S_1\approx 670$;
(2) an intermediate power law 
$F(S)\propto S^{a_2}$, with $a_2\approx -0.165$, and (3) an exponential tail 
$F(S)=p_1 \exp(-S/S_1)$ with $p_1\approx 0.204$ and $S_2=6.97 \times 10^5$.
The size distribution $p(S)$ is, therefore, initially approximately
a power law with  negative exponent $a_1-1\approx -0.29$, followed by a power law with 
negative
exponent $a=a_2-1\approx -1.165$. The latter one extends over approximately
two size decades and ends with an exponential tail characterized by a cutoff
size, $S_1$. The corresponding survival probability $F(T)$ for the avalanche durations 
$T$ is shown in Fig. \ref{Fig6}, b. 
It can be well fitted  by a sum of two exponentials, 
\begin{eqnarray}\label{bi}
F(T)=p_0\exp(-T/T_0)+(1-p_0)\exp(-T/T_1)\;. 
\end{eqnarray}
However, it also seems  to
display an intermediate
power law over about one time decade, from 1 to 10 ms, with the power exponent $b_1\approx -0.523$,
and the cutoff time $T_1\approx 10.93$ ms.
Hence, the probability distribution $p(T)=-dF(T)/dT$ also appears to reflect an intermediate
power law $p(T) \propto T^{b}$ with $b=b_1-1\approx -1.523$. Interestingly, the duration
of avalanches in experiments with organotypic cortical neuronal systems has 
a similar cutoff time of about 10-20 ms, with a maximal avalanche 
duration of about 40-80 ms, which is restricted by the period of 
$\gamma-$oscillations \cite{Plentz12}. The intermediate power law also extends
over about one time decade in the experiments. However, the experimental power law exponent is different, 
$b_{\rm exp}\approx -2$. It should be mentioned in this respect that
 the time bin in the experiments is also very different,
$\Delta t\sim 1-4$ ms. One electrode measures in experiments a contribution of many neurons.
In fact, coarse graining over some unknown $\Delta L$ should be done. The experimental
size exponent $a$ is also different, $a_{\rm exp}\approx -1.5$. It is not, however, the
goal of this paper to provide a model fully consistent with the experimental observations,
which are subject of ongoing research work and some controversy in the literature \cite{Touboul}.
In this respect, a bi-exponential dependence can be perceived as a power law over one intermediate time
decade, as our fit also shows. 

As an additional characteristics of the avalanches size, 
one can also consider  the maximal number of neurons activated at once during an avalanche, 
or the avalanche peak with a
distribution density $p_{\rm max}(S)=-d F_{\rm max}(S)/dS$. The corresponding survival probability,
$F_{\rm max}(S)$, also exhibits a power law, $F_{\rm max}(S)\propto S^{a_2'}$, with $a_2'
\approx -0.175$, in Fig. \ref{Fig6}, c. Hence, $p_{\rm max}(S)\propto S^{a'}$, with 
$a'=a_2'-1\approx -1.175$, which only slightly differs from $a\approx -1.165$ meaning that the avalanche size
is roughly proportional to its peak. However, the cutoff of $F_{\rm max}(S)$ is
super-exponentially sharp, because the maximal number of neurons involved in an avalanche at the same time
is restricted by the total number of neurons in the network.  
Furthermore, $F_{\rm max}$ reveals 
a very large portion of avalanches whose
peak does not exceed 10, which explains the initial stretched exponential 
dependence in Fig. \ref{Fig6}, a.
Strictly speaking, this part of the size distribution (with $L_{\rm thr}=0$) 
reflects a background or basal
 noise, where neurons practically
do not interact with each other, and there are no avalanches of spontaneously
increased activity, which are characterized by a power law distribution.

\begin{figure}
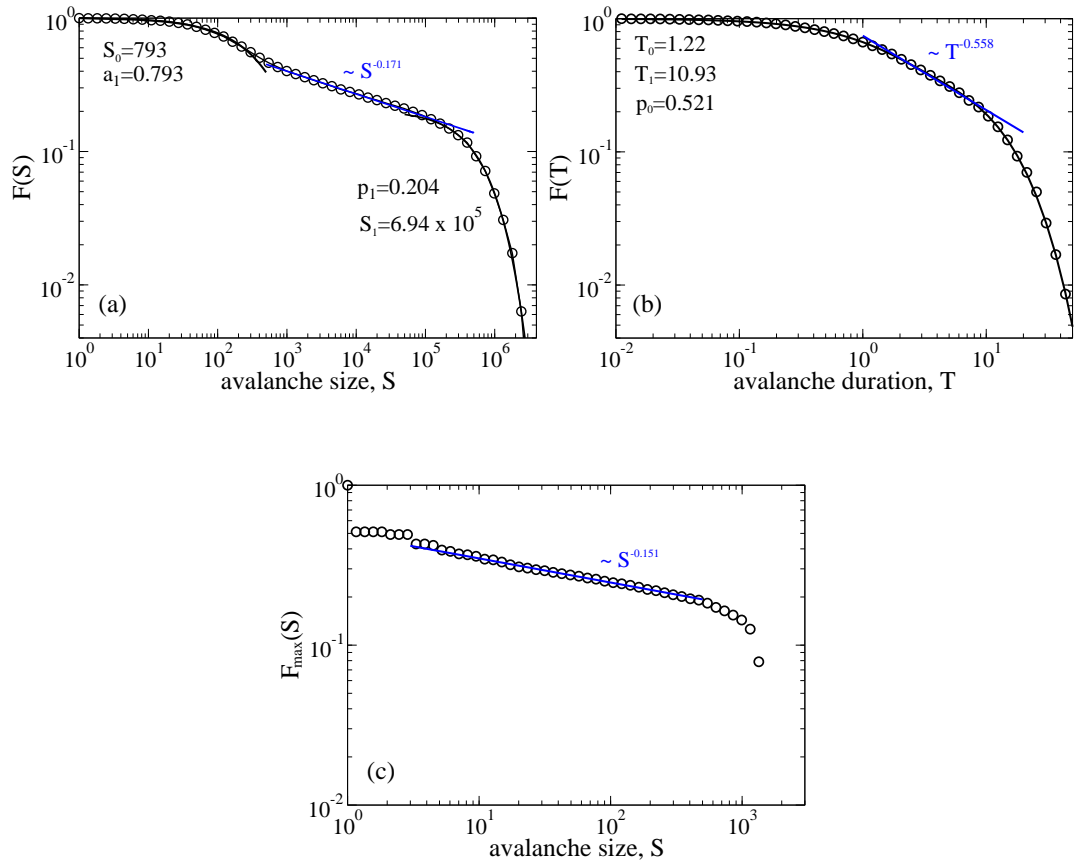

 \centering
 \vspace{0.5cm}
 \includegraphics[width=7cm]{Fig7a.eps} 
 \includegraphics[width=7cm]{Fig7b.eps}
  \vfill \vspace{1cm}
 \includegraphics[width=7cm]{Fig7c.eps}

 \caption{Survival probabilities for the avalanche size (a), duration (b), and peak (c)
 distributions for the network with a time delay, $\gamma_i=\gamma_e=10$.
 Other parameters are the same as in Fig. \ref{Fig6}. The cutoff size of avalanches $S_1$ becomes slightly
  smaller than in Fig. \ref{Fig6}, and the characteristic power law exponents are also slightly changed.
 }
 \label{Fig7} 
\end{figure}

\begin{figure}
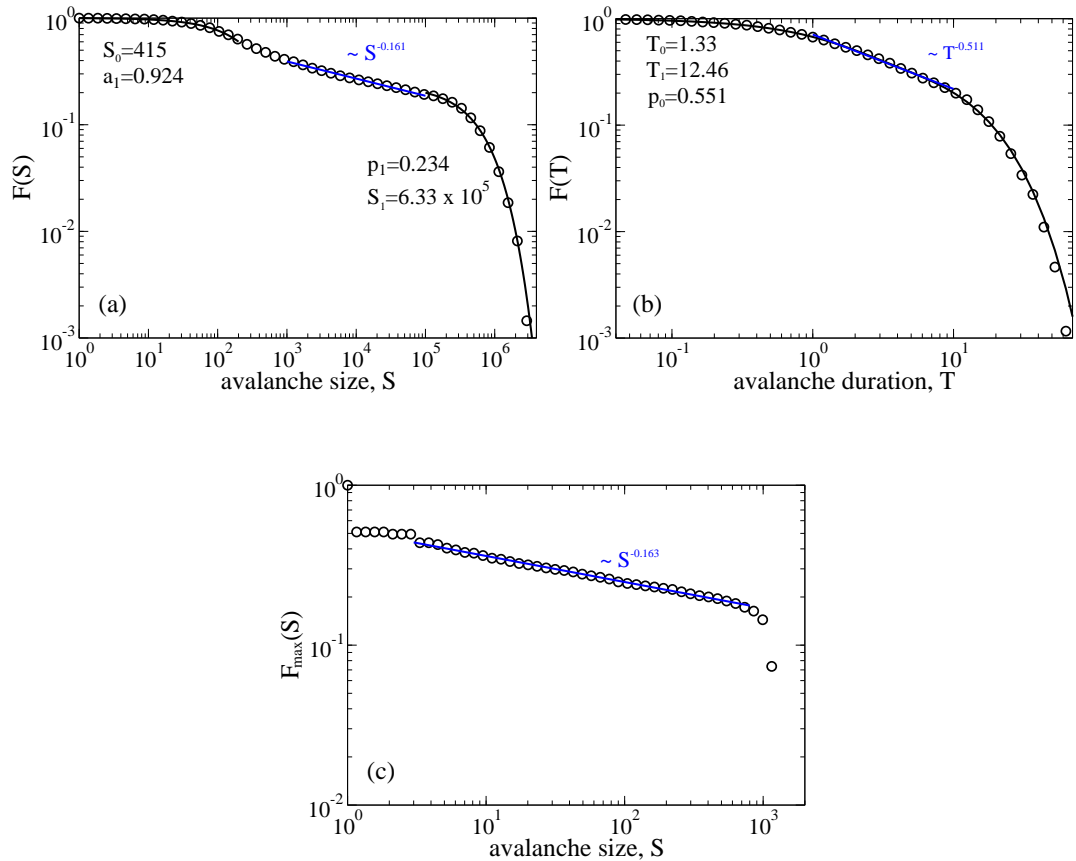

 \centering
 \vspace{0.5cm}
 \includegraphics[width=7cm]{Fig8a.eps} 
 \includegraphics[width=7cm]{Fig8b.eps}
  \vfill \vspace{1cm}
 \includegraphics[width=7cm]{Fig8c.eps}

 \caption{The time delay is increased further with respect
 to the one in Fig. \ref{Fig7}. Here, $\gamma_i=\gamma_e=1$. The other parameters remain the same. 
 The cutoff size of avalanches, 
 $S_1=6.33 \times 10^5$, is now visibly smaller than one without delay, 
 $S_1=6.97 \times 10^5$ in Fig. \ref{Fig6}. The cutoff time $T_1=12.46$ is increased
 with respect to $T_1=10.93$ in Figs. \ref{Fig6}, \ref{Fig7}, i.e. the avalanches last longer.
 The power law exponents here deviate slightly in the opposite direction from the one
 in Fig. \ref{Fig7}. They become closer to the case without delay in Fig. \ref{Fig6}.
 This indicates that the time delay does not affect significantly the power law exponents. 
 }
 \label{Fig8} 
\end{figure}

\begin{figure}
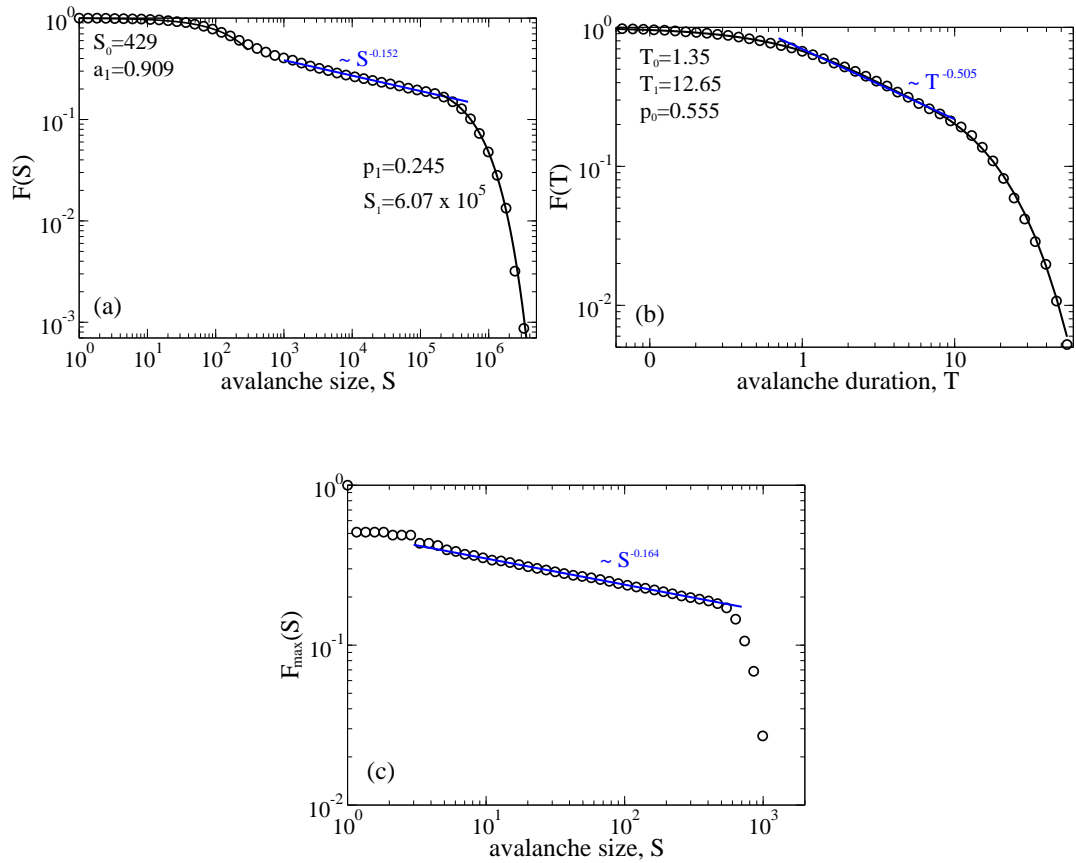

 \centering
 \vspace{0.5cm}
 \includegraphics[width=7cm]{Fig9a.eps} 
 \includegraphics[width=7cm]{Fig9b.eps}
  \vfill \vspace{1cm}
 \includegraphics[width=7cm]{Fig9c.eps}
 \caption{Influence of a further increase of the time delay by an order of magnitude, 
 $\tau_d=10$, on the distributions depicted in Figs. \ref{Fig7}, \ref{Fig8}. Here, $\gamma_i=\gamma_e=0.1$,
 and the other parameters are not changed. $S_1$ drops further to $S_1=6.07 \times 10^5$, and
 $T_1$ slightly increases to $T_1=12.65$. The power law exponents exhibit, however, merely some fluctuations
 without any systematic trend in Figs. \ref{Fig6}-\ref{Fig9}.  
 }
 \label{Fig9} 
\end{figure}

\begin{figure}
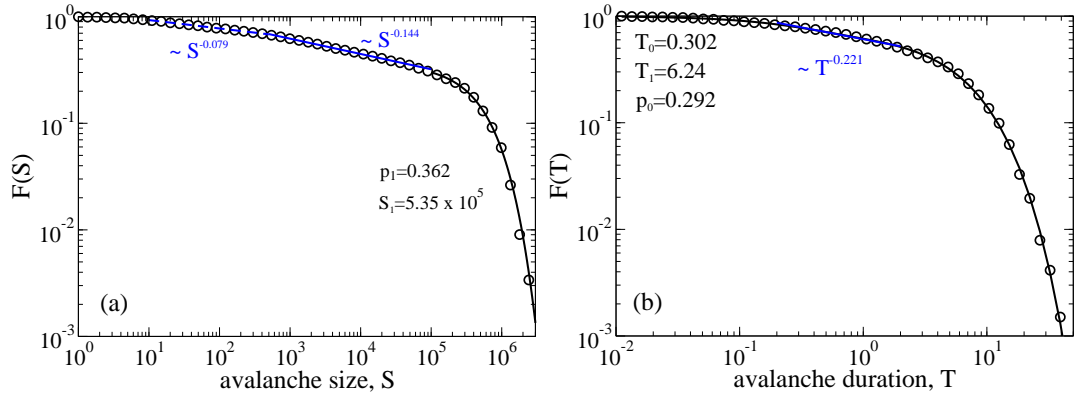

 \centering
 \vspace{0.5cm}
 \includegraphics[width=7cm]{Fig10a.eps} 
 \includegraphics[width=7cm]{Fig10b.eps}
 \caption{Influence of the choice of the detection threshold $L_{\rm thr}$ on the distributions
 of avalanche sizes (a) and time durations (b). Here, $L_{\rm thr}=10$  is used for the data analysis 
 instead of $L_{\rm thr}=0$ in Fig. \ref{Fig7}, for the same data. 
  Noticeably, the initial stretched exponential part
 of the size distribution in Fig. \ref{Fig7}, a disappears. Instead, there appears 
 initially another power law dependence.
 The intermediate power law exponent $a_2$ is, however, pretty robust, $a_2=-0.144$ here versus 
 $a_2=-0.171$ in Fig. \ref{Fig7}. In contrast with this, the intermediate power law exponent in the
 time distribution is changed dramatically from $b_1=-0.558$ in Fig. \ref{Fig7} to $b_1=-0.221$.
 This fact disproves the hypothesis of an intermediate power law in the time distribution.
 It is clearly  bi-exponential, Eq. (\ref{bi}).
 }
 \label{Fig10} 
\end{figure}

Next, we like to clarify how robust  these features are for networks with a time delay.
For this, we study the influence of the mean delay time by decreasing the rates 
$\gamma_e=\gamma_i$ from $10$  through $1.0$ to $0.1$ in 
Figs. \ref{Fig7}, \ref{Fig8}, \ref{Fig9}, respectively.
The mean delay time increases, accordingly, from $0.1$ through $1.0$ to $10$ ms.
Even though the parameters of the distributions do change, these changes are not dramatical.
In particular, the corresponding critical size exponent $a$ changes from $-1.165$ (no delay), to 
$-1.152$,  $-1.171$ and $-1.161$, respectively. Accordingly, the critical time exponent $b$
changes from $-1.523$ (no delay)  to $-1.505$, $-1.558$, and $-1.511$, respectively.
Such changes are not statistically significant, and one cannot detect any systematic
tendency upon a variation of $\tau_d$.
The point is that these exponents are also changed a bit, if we use e.g. 
$2\langle \Delta t\rangle$, or $3\langle \Delta t\rangle$ for 
the time bin (not shown). They also  depend on the threshold $L_{\rm thr}$.
In this respect, if to change $L_{\rm thr}$ from $0$ to $10$, the initial stretched exponential
part of the size distribution practically disappears. However, there appears an 
initial power law instead, see in 
Fig. \ref{Fig10}. Remarkably, the intermediate power law exponent remains rather robust.
It is changed from $a=-1.171$ in Fig. \ref{Fig7},a  to $a=-1.144$ in Fig. \ref{Fig10}, a. 
This is a small variation.
Notice, however, that the results in Fig. \ref{Fig10}, b in fact reject  the hypothesis
that there is an intermediate power law in the time distribution of avalanches. First,
the power law region  changes from larger to smaller times, and also (more important!) 
the corresponding time exponent
changes from $b_1=-0.558$ in Fig. \ref{Fig7}, b to $b_1=-0.221$ in Fig. \ref{Fig10}, b.
Clearly, such a strong influence of the choice of $L_{\rm thr}$ on the ``power law'' 
exponent $b$ makes it  clear that this is not a power law. In fact, 
the time distribution is clearly biexponential.

\begin{figure}
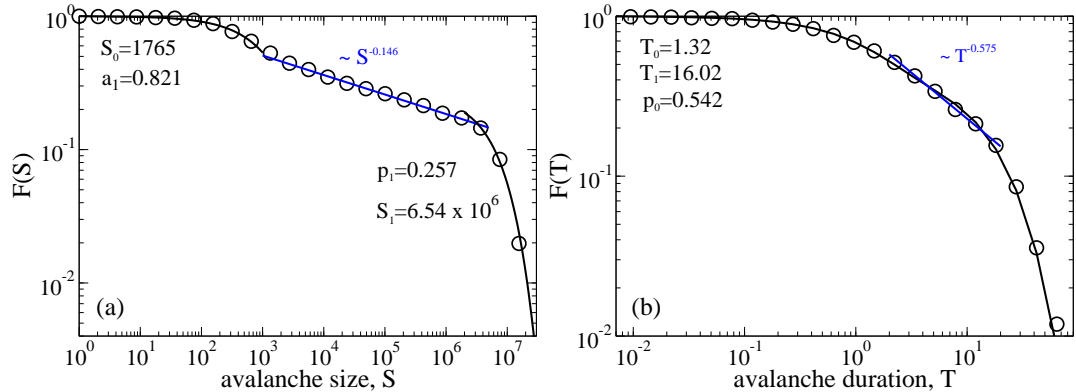

 \centering
 \vspace{0.5cm}
 \includegraphics[width=7cm]{Fig11a.eps} 
 \includegraphics[width=7cm]{Fig11b.eps}
 \caption{Influence of the increased network size on the distributions
 of the avalanche sizes (a) and the time durations (b). Here, $N_i=N_e=10^4$ \textit{vs.} 
 $N_i=N_e=10^3$ in Fig. \ref{Fig9}. The other parameters are the same. The power law regime 
 in the size distribution extends by an order of magnitude, with the cutoff size increased
 to $S_1=6.54\times 10^6$, accordingly. The corresponding power law exponent varies 
 insignificantly. The time cutoff $T_1$ increases in (b) to $T_1=16.02$ from $T_1=12.65$
 in Fig. \ref{Fig9}, i.e. the
 avalanches last longer.
 }
 \label{Fig11} 
\end{figure}

\begin{figure}
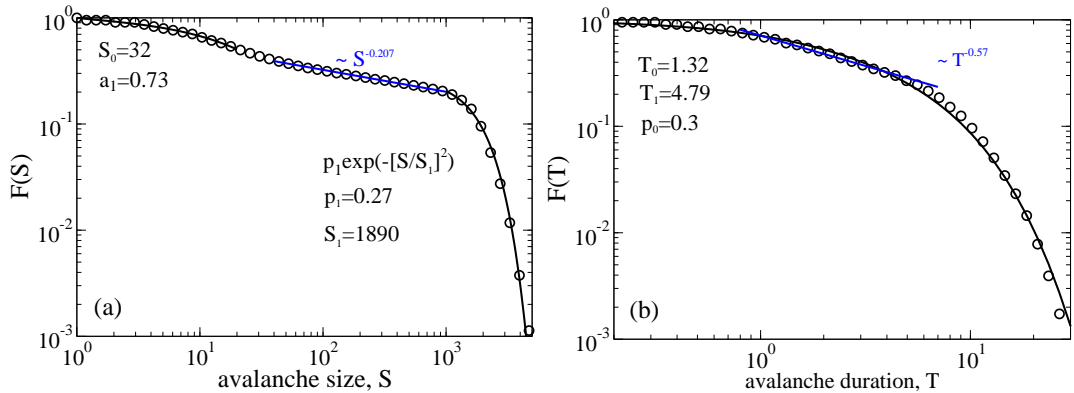

 \centering
 \vspace{0.5cm}
 \includegraphics[width=7cm]{Fig12a.eps} 
 \includegraphics[width=7cm]{Fig12b.eps}
 \caption{Influence of the decreased network size on the distributions
 of avalanche sizes (a) and time durations (b). Here, $N_i=N_e=10^2$ \textit{vs.} 
 $N_i=N_e=10^3$ in Fig. \ref{Fig9}. Other parameters are the same. The power law regime 
 in the size distribution shrinks by an order of magnitude, with the cutoff form
 changed from the exponential in Fig. \ref{Fig9}, a, to the Gaussian here. The intermediate 
 power law exponent is not changed, however, strongly. The time cutoff $T_1$ decreases in (b) 
 to $T_1=4.79$ from $T_1=12.65$
 in Fig. \ref{Fig9}, b, i.e. the avalanches become significantly shorter.
 }
 \label{Fig12} 
\end{figure}

Though plausible until this point, it remains, however, strictly speaking, still  not quite clear if $a$
is  indeed a critical exponent. If true, the extension of the power law
domain of the whole size-distribution and the cutoff size $S_1$ 
should increase with 
the system size accordingly. Indeed, if we increase
the system size tenfold keeping the other parameters the same as in Fig. \ref{Fig9}, the power
law domain in the size distributions also increases by an order of size magnitude,
see in  Fig. \ref{Fig11}, a. The time cutoff $T_1$ also increases in Fig. \ref{Fig11}, b,
i.e. the avalanches become longer. 
Also with decreasing the system size tenfold
the power law domain shrinks accordingly in size, see Fig. \ref{Fig12}, a,
 and the avalanches become essentially shorter, as indicated by the 
decreased cutoff time $T_1$ in Fig. \ref{Fig12}, b.  Such scaling dependencies on the system
size are typical in experiments. 
From this we can conclude that the size exponent $a$ is indeed a critical exponent.
However, within
the considered model the avalanches do gradually vanish with  an increase of the system size.
Therefore, the adjective ``critical'' should  be used also with respect to the exponent $a$
with some reservations. We consider a rather atypical SOC model, even though
the exponent $a$ is by chance close to that of the sandpile model \cite{SOC,Laszlo}. It should also 
be noticed that the initial distributions of the sizes and times and the tail functional 
dependencies can be sensitive to both the system size and the choice of the threshold $L_{\rm thr}$. 
For example, the size distribution exhibits
a Gaussian tail in  Fig. \ref{Fig12}, a, for a small system size. The intermediate power law in
the size distribution is, however, rather
robust, with $a$ being in the range $[-1.207,-1.144]$ for the data presented, with the 
average $\langle a\rangle\approx -1.164$.

\subsubsection{Langevin dynamics of avalanches.}
\begin{figure}
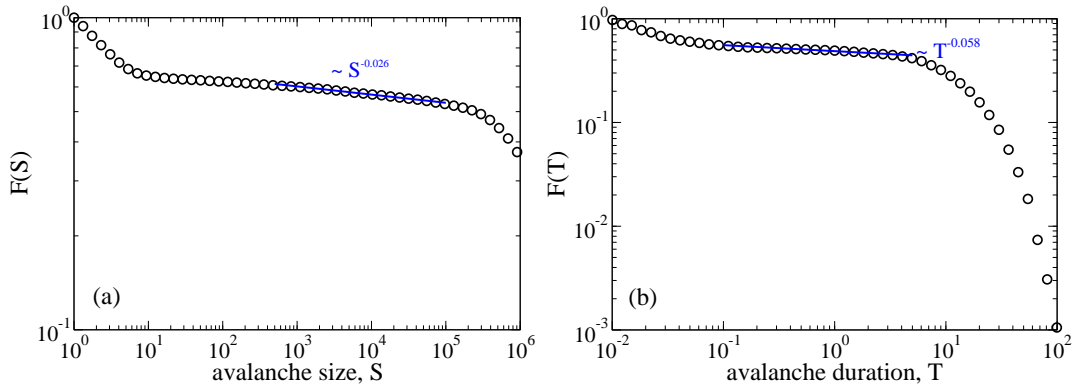

 \centering
 \vspace{0.5cm}
 \includegraphics[width=7cm]{Fig13a.eps} 
 \includegraphics[width=7cm]{Fig13b.eps}
 \caption{The results derived from the Langevin dynamics simulations for the size (a) and time (b)
  distributions, at 
 the same parameters 
 as in Fig. \ref{Fig10}, obtained with $L_{\rm thr}=10$. Notice the dramatical changes.}
 \label{Fig13} 
\end{figure}

Within the Langevin approximation of the discrete state dynamics,
the avalanches look very similar to the ones depicted in Fig. \ref{Fig5}. However,
their statistics is very different. We performed the corresponding Langevin simulations
for the same parameters as in Figs. \ref{Fig7}, \ref{Fig10} with the time step  
taken to be the same $\langle \Delta t\rangle =0.00618608$ as 
 the time bin used to
produce the results in Figs. \ref{Fig7}, \ref{Fig10}. We 
 also used $L_{\rm thr}=10$ to analyze the data, as 
in Fig. \ref{Fig10}. The results shown in Fig. \ref{Fig13} 
reveal similar intermediate power laws both in the size and the time distributions yielding  
$a_L\approx -1.026$, $b_L\approx -1.058$. However, these results differ essentially from 
the results obtained
within the exact dynamic Monte Carlo  simulations, compare Fig. \ref{Fig13} with Fig. \ref{Fig10}. 
This indicates that the Gauss-Langevin or diffusional 
approximation of the genuine discrete state 
dynamics can deliver incorrect results for the fluctuation-induced avalanche dynamics.
This fact makes any analytical theory for the numerical results presented in this work especially
challenging. It is almost hopeless to develop such a theory for the discrete state 
avalanche dynamics within the studied model. Multi-dimensional birth-and-death
processes are very difficult for any analytical treatment.
Within the Langevin  dynamics approximation, or the equivalent multi-dimensional 
Fokker-Planck equation description an analytical treatment is more feasible. 
However, such a theory will not help to understand the critical 
features of the discrete state dynamics, as our numerical results imply.

\section{Summary and Conclusions}

%\begin{figure}
% \centering
% \vspace{0.5cm}
% \includegraphics[width=8cm]{Fig14.eps} 
 
% \caption{Weighted sum of two exponentials, or a power law? Bi-exponential dependence (\ref{bi})
% can look as a power law $F(T)\propto 1/T$ over one time decade.
% }
% \label{Fig14} 
%\end{figure}

In this paper, we studied a generalization of the stochastic 
Wilson-Cowan model of neuronal network
dynamics aimed to incorporate a refractory period delay on the level of individual
elements. Considered as stochastic bistable elements such model neurons exhibit non-Markovian dynamics with
memory, which can be characterized by a non-exponential 
residence time distribution in the resting  state of the neuron (semi-Markovian description), 
or, alternatively, by the related memory
kernel within a generalized master equation description. Such a non-Markovian description
generally allows for a Markovian embedding by enlarging the dynamical space upon introduction of new state
variables.
The simplest two-state non-Markovian model with an exponentially 
decaying memory kernel 
can be embedded as a three state cyclic Markovian model,  where the refractory period
 is exponentially distributed. Multi-state Markovian embedding also allows  
 to treat a special Erlangian distribution of the
refractory periods, which can be sharply peaked at a characteristic delay time. Moreover, models
of bursting neurons and neurons with a power law distributed memory can, in principle,
be considered in this generic
Markovian embedding setup. The approach of Markovian embedding is especially suitable
to treat the mean-field dynamics of the network, which  presents a 
Markovian renewal process in the enlarged space of collective network variables. 
This is the simplest kind of network, where all the elements are virtually connected
in all-to-all fashion.
In this respect, the mean field dynamics of a network of non-Markovian
renewal elements does not represent a renewal process  in the  reduced 
space of  non-Markovian collective
variables. Then, all the elements must be treated individually, keeping trace of their
individual memory. The methodology of Markovian embedding allows to circumvent this
problem for the mean-field dynamics.

In the Wilson-Cowan model, two different sorts of interacting neurons 
are considered, excitatory
and inhibitory. We focused on the simplest non-Markovian generalization of this model, 
where the observed two-state non-Markovian dynamics of a single neuron is 
embedded as a three state cyclic Markovian process. The corresponding nonlinear mean-field dynamics
is four dimensional. It has two dimensions more than in the original model. Moreover, 
it is stochastic and includes mesoscopic fluctuations due to a finite network size. 
 For a sufficiently large system size, stochastic dynamics can be described within
a Langevin equation approximation following the so-called chemical Langevin equation approach,
with the noise terms vanishing in the deterministic limit of infinite size. 
We also exactly simulated
the underlying dynamics as a continuous time Markovian random walk on a 
four-dimensional lattice
using the well-known dynamical Monte Carlo (Gillespie) algorithm. The results of both stochastic approaches agree
well with the deterministic dynamics within an oscillatory regime for a very large number of elements
(several millions). Here, we showed that non-Markovian effects can be very essential. In particular,
even deterministic dynamics with an exponentially decaying memory 
in the space of observable variables can be very different from the dynamics
obtained within the Markovian approximation utilizing a delay-renormalized transfer function -- 
the simplest approach to account for the delay effects. However, already the simplest 
approach allows to describe a dynamical phase transition from the silent network to coherent 
nonlinear oscillations of synchronized neurons upon a change of the delay period.
This important feature is absent in the original Wilson-Cowan model.

In more detail, we investigated the 
avalanche dynamics in a critically balanced network, where the processes
of excitation and inhibition nearly compensate each other in the deterministic limit, where no avalanches
are possible within the model considered. Mesoscopic noise fluctuations make, however, avalanches
possible even in large networks with millions of neurons, where the deterministic description becomes
completely inadequate, very differently from the oscillatory dynamics in such large networks.
This result goes beyond the results in Ref. \cite{Benayon}, where the
avalanches cease to exist for already several tens of thousands elements.
Even though a large delay should suppress avalanches by a transfer function
renormalization if to think within the Markovian approximation, in reality the suppression is much weaker.
 Moreover, it turns out that the power law characterizing the distribution of  
avalanche sizes is very robust with respect to variation of both the delay period, and the system size,
over several orders of magnitude,  as well as the choice of the avalanche threshold. 
The latter fact proves that this is a real power law 
originated due to critical dynamics. It is characterized by a power-law exponent around $a\sim -1.16$
which is similar to the size exponent of the critical sandpile dynamics (though the both models are not really 
comparable). However, it is different from the critical exponent $-1.62$ found
in Ref.\cite{Benayon}, though for different network parameters. 
The distribution of the  avalanches durations is, however, biexponential. We disproved
that it presents a power law within our model, even though it can look like  a power law  over one
 time decade, as
in experiments. In this respect, experiments \cite{Beggs,Plentz12,Beggs08} seem to reveal a 
real power law 
with the critical size 
exponent $a_{\rm exp}\sim -1.5$  because its range extends with the growing system size. However,
the experimental power law in the time duration does not show this important property. As a matter
of fact, it extends over merely one time decade being restricted by the period of $\gamma-$oscillations.
Any power law extending over one time or spatial decade can be fitted by a sum of just 
two exponentials, as we  also show in this work for the time distribution. 
A further research is,
therefore, required to clarify the nature of the apparent power law feature in the avalanche time
distribution for 
the observed neuronal avalanches.

%For example, a bi-exponential (\ref{bi}) with $p_0=0.8$, $T_0=1$ and $T_1=7$
%looks as a power law with $b_1=-1$ yielding power law exponent $b=-2$ between $1$ and $10$, 
%see in Fig. \ref{Fig14}. It is not clear therefore that the time distribution of neuronal avalanches
%presents indeed a  power law dependence in experiments. 
Also very important is that
the Langevin or diffusional approximation does change the critical exponents of the studied avalanche 
dynamics.
There appears a power law in the time distribution, which is absent in the exact simulations,
with the critical Langevin exponent $b_L\sim -1.06$. Also the critical Langevin 
 size exponent is different, $a_L\sim -1.03$. 
This feature should be kept in mind while doing diffusional  approximations 
in other models of critical dynamics. It may deliver incorrect results even for a large number
of elements.

We believe that the results of this work have methodological value and can be extended onto
the dynamics of other networks with delay. They can serve also as a basis for further investigations
of the role of non-Markovian memory effects in the dynamics of Wilson-Cowan type neuronal networks,
including networks of bursting neurons, and networks with nontrivial
topology, which we plan to investigate in future. 

\section{Acknowledgments}

Financial support  by the Deutsche Forschungsgemeinschaft, Grant
GO 2052/1-2 is gratefully acknowledged.

\end{document}